\documentclass[10pt]{elsart}

\usepackage[T1]{fontenc}

\def\dd{\partial}

\begin{document}

\runauthor{Rocha Filho \& Figueiredo}
\begin{frontmatter}
\title{[SADE] A Maple package for the Symmetry Analysis of Differential Equations}
\author[Rocha]{Tarc\'\i sio M.\ Rocha Filho\thanksref{X}}
\author[Rocha]{Annibal Figueiredo}

\address[Rocha]{Instituto de F\'\i{}sica and International Center for Condensed Matter Physics\\ Universidade de
Bras\'\i{}lia, CP: 04455, 70919-970 - Bras\'\i{}lia, Brazil}
\thanks[X]{Corresponding author, e-mail: marciano@fis.unb.br}
\begin{abstract}
We present the package SADE (Symmetry Analysis of Differential Equations) for the determination of
symmetries and related properties of systems of differential equations.
The main methods implemented are: Lie, non classical, Lie-B\"acklund and potential symmetries, invariant solutions,
first-integrals, N\"other theorem for both discrete and continuous systems,
solution of ordinary differential equations, order and dimension reductions using Lie symmetries,
classification of differential equations, Casimir invariants, and the
quasi-polynomial formalism for ODE's (previously implemented by the authors in the package QPSI) for the determination 
of quasi-polynomial first-integrals, Lie symmetries and invariant surfaces. Examples of use of the package are given.
\end{abstract}
\end{frontmatter}

%\begin{titlepage}

%\maketitle

\newpage

%\baselineskip19truept

{\large \bf PROGRAM SUMMARY} 

\vspace{5mm}
\noindent \em Title of the program: \bf SADE

\noindent \em Catalogue identifier: \rm None

\noindent \em Program obtainable from: \rm The author by e-mail.

\noindent \em Operating systems under which the program has been tested: \rm UNIX/LINUX systems and WINDOWS

\noindent \em Programming language used: \rm MAPLE 13 and MAPLE 14

\noindent \em No. of bytes in a word: \rm 32

\noindent \em No. of bytes in distributed program: \rm 300 KB

\noindent \em Distribution format: \rm zip or gzip

\noindent \em Card punching code: \rm ASCII


\noindent \em Nature of the physical problem: \rm Determination of analytical properties of
systems of differential equations, including symmetry transformations, analytical
solutions and conservation laws.

\noindent \em Method of resolution: \rm The package implements in MAPLE some algorithms
(discussed in the text) for the study of systems of differential equations.

\noindent \em Restrictions on the complexity of the problem: \rm Depends strongly on the system and on the
algorithm required. Typical restrictions are related to the solution of a large over-determined system
of linear or non-linear differential equations.

\noindent \em Typical running time: \rm Depends strongly on the order, the complexity of the differential
system and the object computed. Ranges from seconds to hours.

%\end{titlepage}

%\baselineskip19truept

{\large {\bf LONG WRITE-UP}}

\section{Introduction}

Natural phenomena are very often modeled by differential equations, which exhibit a plethora of
dynamical behaviors. These can be classified somewhat vaguely in two categories: regular and irregular,
according to the complexity exhibited by its solutions. The notion of integrability is then used as an attempt
to put a more stringent distinction between {\it regular} and {\it irregular} behavior.
Usually a regular behavior is characterized by the existence of conservation laws that strongly  restrict
the types of solution a system can exhibit. Even for non-integrable systems some of these laws can be obtained.
Also the determination of particular analytical solutions
for both Ordinary (ODE's) and Partial Differential Equations (PDE's) is of utmost importance in many fields of physics and
applied sciences. These solutions are helpful to shed some light and gain insight on the physics of the system,
and are also useful as benchmarks for numerical methods. Almost all known analytical solutions in physics
are solutions invariant under one or more symmetry transformations.

The theory of symmetry
transformations of Differential Equations (DE's) was introduced by Lie in the end of the XIX century~\cite{ibragimov}.
Solutions invariant under symmetry transformations are called invariant solutions, and different methods
are described in the literature (see References \cite{c1,rosenau,vu2,bluman} and references therein).
Olver and Rosenau showed that new solutions can be obtained by requiring that they are invariant under
infinitesimal symmetry transformations while also preserving additional side conditions~\cite{rosenau}.
Non classical symmetries were introduced by Bluman and Cole and are based on the idea that the required analytical
solution is invariant under symmetry transformations preserving both the form of the differential equation and
the invariant solution condition~\cite{bluman}. This approach is less restrictive in the sense that there exist
usually more non classical symmetries than Lie (classical) symmetries, the latter being a subset of the former.
Other generalizations of the classical Lie method considered here are
potential symmetries~\cite{kumei} and Lie-B\"acklund transformations~\cite{anderson}.
For a first introduction to Lie Symmetries see~\cite{hydon,steeb}, and~\cite{olver,c1} for a more complete and formal
approach. A description of methods for solving differential equations using Symmetries is found in~\cite{stephani}.

Different packages in computer algebra systems exist implementing Lie symmetry computations and related methods:
SPDE by Schwarz~\cite{schwarz}, CRACK and LIEPDE by Wolf~\cite{wolf1,wolf2} and DIMSYM by
Sherring and Prince~\cite{sherring} in REDUCE,
LIE and BIGLIE by Head~\cite{head1,head2} in MUMATH and MATHLIE by Baumann~\cite{baumann} in Mathematica.
For MAPLE there are also some useful
packages: PDEtools by Cheb-Terrab~\cite{cheb} which is distributed since Release 11,
DESOLV by Vu and Carminati~\cite{carminati,vu2},
and GeM by Cheviakov~\cite{cheviakov}. For good reviews
with a comparison between some of these packages see References~\cite{butcher} and~\cite{hereman}.
The package QPSI by the authors implements the Quasi-Polynomial formalism for
symmetry generators, first-integrals and invariant hyper-surfaces for ODE's~\cite{qpsi}, now part
of the present package.

In this work we present the package {\it Symmetry Analysis of Differential Equations} (SADE) in MAPLE,
for the computation of Lie, Lie-B\"acklund and nonclassical symmetries, invariant solutions,
first-integrals, N\"other theorem for both discrete and continuous systems,
quasi-polynomial first-integrals and symmetry generators,
solution and reduction of order or dimension for ODE's,
classification of differential equations, invariant surfaces and Casimir invariants~\cite{hydon,olver,nos1,qpsi},
and some other features presented below.
Our package is well suited for efficiently computing Lie symmetries of large systems,
as for instance the Yang-Mills with $SU(2)$ and $SU(3)$ gauge group~\cite{weinberg}, and
has been used in the last years in our group in different
applications~\cite{ruben,erica,cardeal,lyap,liu}.

Our aim in developing this package was to implement these methods, being
as user friendly as possible for researchers in many fields of pure and applied sciences,
and still being capable to handle reasonably complicated systems of equations.
The paper is structure in the following way:
in section 2 we briefly revise the mathematical methods implemented. A discussion of the
heuristics for the solution of the determining system for Lie and nonclassical symmetries
of linear and non-linear overdetermined systems of PDE's
is given in section 3.
The package routines are described in section 4, and
some illustrative examples are given in section 5. Benchmarks for computing Lie symmetries are given
in section 6. We close the paper with some concluding remarks in section 7.

\section{Methods}

\subsection{Lie symmetries of differential equations}

Let $\{u_1,\ldots,u_n\}\equiv u$ be a set of functions (dependent variables) of the (independent) variables
$\{ x_1,\ldots,x_m\}\equiv x$. A system of $p$ differential equations
satisfied by the $n$ functions $u_j(x)$ can be written as
\begin{equation}
F_\mu(u_j,x_i,u_{jI})=0,\hspace{3mm}\mu=1,\ldots,p,
\label{eq1}
\end{equation}
with
\begin{equation}
u_{jI}\equiv u_{j,i_1,\ldots,i_k}=\partial^k u_j/\partial x_{i_1}\cdots\partial x_{i_k},\hspace{3mm}I\equiv i_1,\ldots,i_k.
\label{jetnot}
\end{equation}
For $m=1$ eq.~(\ref{eq1}) is a set of ODE's.

A transformation of variables
\begin{eqnarray}
 & & x_i'=x'_i(u,x),\nonumber\\
 & & u'_j=u'_j(u,x),
\label{transfin}
\end{eqnarray}
is a symmetry transformation of eq.~(\ref{eq1}) if
\begin{equation}
F_\mu(u'_j,x'_i,u'_{jI})=0,
\label{eq3}
\end{equation}
where $u'_{jI}\equiv\partial^k u'_j/\partial x'_{i_1}\cdots\partial x'_{i_k}$,
whenever eq.~(\ref{eq1}) holds, i.~e.~if eq.\ (\ref{eq1}) is form invariant under (\ref{transfin}), or equivalently,
if eq.~(\ref{transfin}) maps a solution into another solution of eq.~(\ref{eq1}). Such transformations are called
Lie symmetries (point symmetries). The set of Lie symmetries of a (system of) differential equation(s) is a Lie group,
and therefore can be obtained from the knowledge of the infinitesimal transformations (in fact only the subgroup
of transformations connected to the identity transformation)~\cite{olver}. The infinitesimal symmetries can be written as:
\begin{eqnarray}
 & & x_i'=x_i+\epsilon\theta_i(u,x),\nonumber\\
 & & u'_j=u_j+\epsilon\eta_j(u,x),
\label{eq2}
\end{eqnarray}
where $\epsilon$ is an infinitesimal parameter and $\theta_i$ and $\eta_j$ are functions of the
dependent and independent variables.
The infinitesimal symmetry generator of  transformation (\ref{eq2}) is
\begin{equation}
{\bf G}=\sum_{j=1}^n \eta_j\frac{\partial}{\partial u_j}+
\sum_{i=1}^m \theta_i\frac{\partial}{\partial x_i}.
\label{eq4}
\end{equation}
The set of all infinitesimal symmetry generators form a Lie algebra with respect to the commutation operation.

In order to determine the invariance condition of~(\ref{eq1}) under the infinitesimal transformation~(\ref{eq2}), we
note that
\begin{equation}
\frac{\partial u'_j }{\partial x'_i }=\frac{\partial u_j}{\partial x_i}+\epsilon\left[
\frac{\partial\eta_j}{\partial x_i}-\sum_{l=1}^m
\frac{\partial u_j}{\partial x_l}\frac{\partial\theta_l}{\partial x_i}\right]
\equiv\frac{\partial u_j}{\partial x_i}+\epsilon\eta^{(1)}_{ji}.
\label{eq5}
\end{equation}
The transformation rules for higher order derivatives can be obtained similarly. In the general
case we have:
\begin{equation}
\frac{\partial^k u'_j}{\partial x'_{i_1}\cdots\partial x'_{i_k}}=
\frac{\partial^k u_j}{\partial x_{i_1}\cdots\partial x_{i_k}}+\epsilon\eta^{(k)}_{j,i_1\cdots i_k}
\equiv u_{j,i_1\cdots i_k}+\epsilon\eta^{(k)}_{j,i_1\cdots i_k},
\label{eq6}
\end{equation}
with $\eta^{(k)}_{j,i_1\cdots i_K}$ functions of the independent and dependent variables and its derivatives~\cite{olver}.
Supposing that the highest derivative in (\ref{eq1}) is of order $k$ we can express its invariance
under an infinitesimal transformation by
\begin{equation}
{\bf G}^{(k)}F_\mu=0,
\label{eq7}
\end{equation}
where ${\bf G}^{(k)}$ is the $k$-th prolongation of the generator ${\bf G}$ in (\ref{eq4})
and is obtained from eq.~(\ref{eq6}) as:
\begin{equation}
{\bf G}^{(k)}=\sum_{i=1}^m \theta_i\frac{\partial}{\partial x_i}+\sum_{j=1}^n
\eta_j\frac{\partial}{\partial u_j}+\sum_{l=1}^k\sum_{i_1,\ldots,i_l}\eta^{(l)}_{j,i_1\cdots i_l}
\frac{\partial}{\partial u_{j,i_1\cdots i_l}}.
\label{eq8}
\end{equation}
Using the orthonomic form of eq.~(\ref{eq1}) to eliminate highest order derivatives from eq.~(\ref{eq7}), and equating to zero
the coefficients of the remaining derivatives, or more precisely the coefficients of linearly independent functions of the latter,
we obtain the determining system for the symmetry transformations of eq.~(\ref{eq1}).
The reduction to the orthonomic form of eq.~(\ref{eq1}) is performed using standard methods (see~\cite{rif,rif2} and references therein).
If the reduction is not possible SADE will issue an error message before aborting the calculations.

The symmetry generator in eq.~(\ref{eq4}) is equivalent to the following
evolutionary form:
\begin{equation}
{\bf \tilde{G}}=\sum_{j=1}^n\left[\eta_j-
\sum_{i=1}^m \frac{\partial u_j}{\partial x_i}\theta_i\right]\frac{\partial}{\partial u_j}
\equiv\sum_{j=1}^n Q_j\frac{\partial}{\partial u_j}.
\label{canform}
\end{equation}
Both forms as given in eqs.~(\ref{eq4}) and~(\ref{canform}) describe the same transformation,
in the sense that they map a given solution to the same transformed solution.

\subsubsection{Quasi-Polynomial Symmetries}
\label{invten}

First order differential equations usually admit an infinite dimensional Lie symmetry
group. To determine their Lie symmetries it is usually necessary to impose an ansatz
on its symmetry generators. One possibility is to suppose that the coefficients $\eta_i$ of the
symmetry generator are polynomial functions of the dependent variables. A more general ansatz consists to consider the class of
quasi-polynomial functions, as introduced in ref.~\cite{nos1}, previously implemented by
the authors in the package QPSI~\cite{qpsi}, and now included in SADE.

A system of ODE's of the form
\begin{equation}
{\dot{x}}_i=x_i\sum_{j=1}^mA_{ij}\prod_{k=1}^n{x_k}^{B_jk}\,;\;\;\;i=1,...,n%
\,,  \label{eq1b}
\end{equation}
is called Quasi-Polynomial (QP)~\cite{11}. In eq.~(\ref{eq1b})
$A_{ij}$ and $B_{jk}$ are real or complex constants and $m$
is the number of different quasi-monomials in (\ref{eq1b}). We define a new
set of variables by the Quasi-Monomial Transformation (QMT):
\begin{equation}
y_i=\prod_{k=1}^n{x_k}^{{\stackrel{\sim }{B}}_{ik}},
\label{eq2b}
\end{equation}
where ${\stackrel{\sim }{B}}_{ik}=B_{ik}$ for all $i$ and $k\leq n$, ${%
\stackrel{\sim }{B}}_{ik}=0$ for $i\leq n$ and $k\geq m$, and ${\stackrel{%
\sim }{B}}_{ik}=\delta _{ik}$ for $n\leq i,k\leq m$, in such way that the
inverse transformation is also a QMT with the exponent matrix given by ${%
\stackrel{\sim }{B}}^{-1}$ (the case with $\stackrel{\sim}{B}$ singular can also be handled as
discussed in~\cite{11}).
System (\ref{eq1}) is cast by transformation (\ref{eq2b})
into a quadratic system of equations, the Lotka-Volterra form:
\begin{equation}
{\dot{y}}_i=y_i\sum_{j=1}^mM_{ij}y_j\,,\;\;\;i=1,...,m\,,
\label{eq3b}
\end{equation}
where the matrix $M$ is given by:
\begin{equation}
M=BA\,.  \label{eq4b}
\end{equation}
Lotka-Volterra equations are extensively studied in the literature. Many results obtained for this special class of
equations can then be recast into the more general QP form~\cite{nos1,ruben,lyap,11,nos2,markus}.

The central result obtained in Ref.~\cite{nos1} is that any quasi-polynomial
symmetry generator ${\bf G}$, such that $\theta_i=0$ and $\eta_i$ a quasi-polynomial function of the dependent variables,
can be decomposed as:
\begin{equation}
{\bf G}=\sum_{i}{\bf G}^{(i)}\,,
\label{eq16it}
\end{equation}
with:
\begin{equation}
{\bf G}^{(i)}=y^{\xi^{(i)}}{\bf T}^{(i)}\, , \;\;\; [{\bf F},{\bf G}^{(i)}]=0\, ,
\label{eq17it}
\end{equation}
where ${\bf F}$ is the flow associated to the quasi-polynomial system~(\ref{eq1b}):
\begin{equation}
{\bf F}\equiv\sum_{i,j=1}^m M_{ij} y_i y_j\frac{\partial}{\partial y_i},
\label{lvflow}
\end{equation}
$y^{\xi^{(i)}}$ is a quasi-monomial:
\begin{equation}
y^{\xi^{(i)}}\equiv (y_1)^{\xi_1^{(i)}}\cdots(y_m)^{\xi_m^{(i)}},
\end{equation}
with $\xi_j^{(i)}$ real numbers (see eq.~\ref{eq10} below) and ${\bf T}^{(i)}$ a polynomial
semi-invariant vector field satisfying:
\begin{equation}
[{\bf F},{\bf T}^{(i)}]=\lambda^{(i)}{\bf T}^{(i)}\,.
\label{eq18it}
\end{equation}
The eigenvalue $\lambda^{(i)}$ is a linear function of the form
\begin{equation}
\lambda^{(i)}=\sum_j \lambda^{(i)}_j y_j.
\end{equation}
Both $\lambda^{(i)}_j$ and $\xi^{(i)}$ are solutions of the equation:
\begin{equation}
\sum_j \xi^{(i)}_j M_{jk}=-\lambda^{(i)}_k.
\label{eq10}
\end{equation}

It is straightforward to show that if a symmetry generator (including the flow ${\bf F}$) can be
written as a linear combination of the remaining generators, then the coefficients of the expansion
(as functions of $x_i$) are first-integrals of system~(\ref{eq1b}).

\subsection{Lie-B\"acklund symmetries}

Lie symmetries are diffeomorphisms on the space of dependent $u_j$ and independent $x_i$ variables.
Lie-B\"acklund, or generalized, symmetries depend also on derivatives of $u_j$.
We present here a brief account of how to compute the generators of Lie-B\"acklund symmetries
(for more details see~\cite{olver} and~\cite{anderson}).
The determining equations for Lie-B\"acklund symmetries are more easily obtained using the evolutionary
form~(\ref{canform}), with $\eta_j=\eta_j(u_j,x_i,u_{jI})$ and
$\theta_i=\theta_i(u_j,x_i,u_{jI})$, with prolongation 
\begin{equation}
{\bf G}^{(k)}=\sum_{j=1}^n\sum_I D_I Q_j\frac{\partial}{\partial u_{jI}},
\label{lbcan}
\end{equation}
where $I\equiv i_1,\ldots,i_m$, $D_I\equiv d^k/dx_{i_1}\cdots dx_{i_k}$, $D_0\equiv 1$,
and the summation  over $I$ is taken for all values of indices such that $|I|=i_1+\ldots+i_k\le k$.
The invariance condition can be expressed as
\begin{equation}
{\bf G}^{(k)} F_\mu(x,u,u_I)=0,
\label{lbinvcond}
\end{equation}
with $k$ the maximum differentiation order of $u_j$ in $F_\mu$. In order to equate coefficients
of independent derivatives of $u_j$ in~(\ref{lbinvcond}), we distinguish dependent and independent derivatives
of $u_j$ using the original system in the orthonomic form, and its differential consequences. In this way only
independent derivatives remain, and at this step each coefficient of
derivatives that are not arguments of $Q_j$ is equated to zero, yielding the determining system for
Lie-B\"acklund symmetries. Its solution demands a greater computational effort than the solution of the
analogous determining system for Lie symmetries.

\subsection{Reduction of PDE's and invariant solutions}

Symmetries of a differential system can be used to
construct analytical solutions or a reduction into a system depending on a smaller number of independent variables.
A symmetry generator
as given in~(\ref{eq4}) can be transformed, by a change of dependent $r_i=r_i(u,x)$ and independent
variables $s_j=s_j(u,x)$, called canonical coordinates, into the form
\begin{equation}
{\bf G}_1=\frac{\partial}{\partial s_1}.
\label{eqcoordcan}
\end{equation}
Solutions invariant under the symmetry generated by ${\bf G}_1$ do not depend
on $s_1$. In this way we obtain a reduction into a system with $n-1$ independent variables.
More generally $p<m$ symmetry
generators can de used to reduce to a new system with $m-p$ independent variables, on the condition
that a set of mutual canonical variables exists for the set of $p$ generators. For $p=m-1$
we obtain a system of ODE's. The latter, if solvable, then yields an analytical (particular) solution
for the original system.
In practice, considering $p$ generators ${\bf G}^{(i)}$, $i=1,\ldots,p$, we look
for solutions $u_i(x)$ satisfying
\begin{equation}
{\bf \tilde{G}}^{(i)}u_j(x)=0,\hspace{5mm}j=1,\ldots,n,
\label{solinv}
\end{equation}
where ${\bf \tilde{G}}^{(i)}$ is the evolutionary form~(\ref{canform}).
Equation~(\ref{solinv}) is a linear system usually simpler to solve than the original system using the characteristics method.
Replacing its solution into the original system yields a reduced system with $n-p$ independent variables.

\subsection{Nonclassical symmetries}

Lie symmetries maps the set of all solutions of a differential system into itself. Invariant
solutions then correspond, among all solutions, to those that are invariant under one or more
symmetry transformations. More generally, nonclassical symmetries transform a solution, still to be determined,
into itself.
This amounts to require that both~(\ref{eq1}) and the invariance condition:
\begin{equation}
Q_i={\bf \tilde{G}}\:u_i(x)=\left[\eta_i-
\sum_{j=1}^m \frac{\partial u_i}{\partial x_j}\theta_j\right]=0,
\label{invcond}
\end{equation}
are invariant under~(\ref{eq2}). This is expressed by:
\begin{equation}
{\bf G}^{(k)}F_\mu=0,
\label{eqdetnocl1}
\end{equation}
and
\begin{equation}
 {\bf G}^{(k)}Q_i=0.
\label{eqdetnocl2}
\end{equation}
A more detailed account of nonclassical symmetries is given in Ref.~\cite{boussinesq}.
It can be shown that eq.~(\ref{eqdetnocl2}) holds whenever eq.~(\ref{eqdetnocl1}) is satisfied.
The resulting system~(\ref{eqdetnocl1}) is non-linear in the unknowns $\eta_i$ and $\theta_i$
as it must be solved modulo eq.~(\ref{eqdetnocl2}), and thus much harder to solve
than the linear determining system for Lie symmetries. The set of all nonclassical symmetries
include all Lie symmetries, and do not form a vector space
(no associated Lie algebra). Another useful property is that if ${\bf G}$ is the generator
of a nonclassical symmetry, then $F(u,x)\:{\bf G}$ also generates a nonclassical symmetry,
for any arbitrary (sufficiently differentiable) function $F$. As a consequence and without loss of generality,
we consider the cases with $\theta_1=1$ or $\theta_1=0$. In the later case there are still two possibilities:
either $\theta_2=1$ or $\theta_2=0$, and so on.

Computer algebra determination of the invariance
condition for nonclassical symmetries can lead to infinite loops when replacing dependent derivatives
from eq.~(\ref{eqdetnocl2}) into eq.~(\ref{eqdetnocl1})~\cite{mansfield}.
This is avoided in our approach by the following steps:
\begin{enumerate}
\item Chose an independent variable $x_k$.
\item Solve the invariance conditions~(\ref{invcond}) for all derivatives.\label{passo2}
$\partial u_i/\partial x_k$ , $i=1,\ldots,n$.
\item From the result of the previous step, eliminate all
derivatives with respect to $x_k$ in~(\ref{eq1}).
\item Determine the invariance condition using the resulting differential system.
\item Replace in the determining system all derivatives of $u_i$ with respect to $x_k$ using step~(\ref{passo2}).
\end{enumerate}
The variable $x_k$ is chosen such that $MaxDer(x_k)<MaxDer(x_i)$ for $i\neq k$,
with $MaxDer(x_i)$ the maximum derivative order of any dependent variable with respect to $x_i$ in~(\ref{eq1}).
This usually results in a ``simpler'' determining system.

\subsection{Potential symmetries}

For systems in conserved form, potential symmetries can be used to construct invariant solutions
that are not obtainable neither from Lie nor nonclassical symmetries~\cite{kumei,blumanreid}.
A partial differential equation is said to be in a conserved form if it can be written as:
\begin{equation}
\sum_{i=1}^m\frac{\partial}{\partial x_i}
F_i(u_j,x_i,u_{jI})=0.
\label{potform}
\end{equation}
This implies that there exists $m(m-1)$ components (potentials) $\Psi_{i,j}$ ($i<j$) of an antisymmetric
tensor such that
\begin{equation}
F_i=\sum_{i<j=1}^m (-1)^j\frac{\partial\Psi_{ij}}{\partial x_j}+
\sum_{j<i=1}^m (-1)^{i-1}\frac{\partial\Psi_{ji}}{\partial x_j}.
\label{potsyst}
\end{equation}
The generalization to a system of PDE's is straightforward
(each equation must be put in a conserved form). Eq.~(\ref{potsyst}) is a system of $M$ PDE's with
$1+m(m-1)/2$ dependent variables ($u_j$ and the potentials). For $M\geq 3$ the system is under-determined, and some
``gauge'' conditions on the potentials must be given~\cite{blumanreid}.

An infinitesimal symmetry of eq.~(\ref{potsyst})
\begin{eqnarray}
 & & u'_i=u_i+\epsilon\:\eta_i(u,\Psi,x),\nonumber\\
 & & \Psi'_{ij}=\Psi_{ij}+\epsilon\:\xi_{ij}(u,\Psi,x),\nonumber\\
 & & x'_i=x_i+\epsilon\:\theta_i(u,\Psi,x),
\end{eqnarray}
is a potential symmetry of the original system~(\ref{potform}) if $\eta_i$ or $\theta_i$ depend on $\Psi_{ij}$.
As a result the transformation for $u$ and $x_i$ is non-local, since it depends on the potentials,
which are solutions of eq.~(\ref{potsyst}). A potential symmetry can then be used to
reduce eq.~(\ref{potform}) and, in some cases, to obtain invariant solutions.

\subsection{Symmetries and conservation laws}

\subsubsection{Quasi-polynomial first-integrals}

\label{qpfirst}

For the special case of quasi-polynomial first-order systems, we first obtain the associated
Lotka-Volterra form~(\ref{eq3b}), and define a semi-invariant (Darboux polynomial) as a polynomial function
$f(y)$ such that
\begin{equation}
{\dot f}=\sum_{j=1}^m {\dot y}_i\frac{\partial f}{\partial y_j}=\lambda f\, ,
\end{equation}
where the eigenvalue $\lambda$ is a function of $y_1,\ldots,y_m$.

For the Lotka-Volterra form the following properties were proved in \cite{nos1,nos2}:

\noindent (i) $\lambda$ is a linear function, i.e.,
\begin{equation}  \label{eq5b}
\lambda = \sum_{j=1}^m\lambda_jy_j \, .
\end{equation}

\noindent (ii) Any polynomial semi-invariant $f$ can be decomposed as:
\begin{equation}
f=\sum_{p}f^{(p)}\, ,
\label{eq6b}
\end{equation}
where $f^{(n)}$ is a homogeneous polynomial of degree $p$. Furthermore, each
$f^{(n)}$ is also a semi-invariant with the same eigenvalue as $f$.
Any Quasi-Polynomial first-integral can be decomposed as:
\begin{equation}
{\cal J}=\sum_{p}y^{\xi^{(p)}}f^{(p)}\, ,
\label{eq11b}
\end{equation}
where $y^{\xi^{(p)}}f^{(p)}$ is a first-integral with $\xi^{(p)}$ a solution of eq.~(\ref{eq10}).

If one of the Quasi-Monomials in~(\ref{eq2}) is a constant,
$\lambda$ may also admit a constant value with respect to the original variables $x_k$, and a
first-integral can be obtained by multiplying the corresponding semi-invariant by $\exp(-\lambda t)$.
It is easy to show that for $f^{(1)}$ and $f^{(2)}$ semi-invariants with respective eigenvalues
$\lambda^{(1)}$ and $\lambda^{(2)}$,  $f^{(1)}f^{(2)}$ and $f^{(1)}/f^{(2)}$
are also-semi-invariants with eigenvalues $\lambda^{(1)}+\lambda^{(2)}$ and $\lambda^{(1)}-\lambda^{(2)}$,
respectively.
The first integrals are then obtained by combinations of the form
\begin{equation}
QP_1(x)\left[QP_2(x)\right]^{\pm 1}\exp(\rho t),
\label{invariante1}
\end{equation}
such that it has a vanishing eigenvalue, and therefore has a zero time derivative.
Analogously, it is straightforward to implement the computation of first-integrals of the form
$P_1(x)+\log(x^\xi)$ and $P_2(x,\ln(x))$, where $P_1$ and $P_2$ are polynomials in their arguments.
A similar result also holds for quasi-polynomial symmetries~\cite{qpsi}.

\subsubsection{N\oe ther theorem}

Many systems of interest are described by equations that can be deduced from a variational principle, with
action $S$ defined by
\begin{equation}
S\equiv\int_{\mathcal C} {\mathcal L}(u_j,u_{j,i},x_i)\: d^m x.
\label{eq22}
\end{equation}
where $u_j$ ($j=1,\ldots,n$) are the dependent variables, $x_i$ ($i=1,\ldots,m$) the
independent variables and $u_{j,i}\equiv \partial u_j/\partial x_i$. The lagrangian of the
system is denoted by ${\mathcal L}$ and ${\mathcal C}$ is a bounded region of the $m$-dimensional
space of independent variables. The differential system is obtained from the requirement that
$S$ is an extremum for any solution $u_j(x)$.

N\"other theorem~\cite{c8} states that every symmetry transformation of the action $S$
of the form
\begin{eqnarray}
 & & x'_i=x_i+\epsilon\:\theta_i(u,x),\nonumber\\
 & & u'_j=u_j+\epsilon\:\eta_j(u,x),
\label{eq24}
\end{eqnarray}
is related to a conservation law. The invariance of $S$ under (\ref{eq24}) implies that
\begin{equation}
{\mathcal D L}=\frac{df_i}{dx_i},
\label{eq24bis}
\end{equation}
with
\begin{equation}
{\mathcal D}\equiv \sum_i\theta_i\frac{\partial}{\partial x_i}+
\sum_j u_j\frac{\partial}{\partial u_j}
+\sum_{j,i}\left(\frac{d\eta_j}{dx_i}-\sum_ku_{j,k}\frac{d\theta_k}{dx_i}\right)
\frac{\partial}{\partial u_{j,i}}+\sum_i\frac{d\theta_i}{dx_i},
\label{eq24tris}
\end{equation}
where $f_i$ are also unknowns to be determined from condition (\ref{eq24bis})
alongside with the $\theta_i$'s and the $\eta_\mu$'s. The associated first-integral
or conserved current is given by
\begin{equation}
I_i={\mathcal L}\theta_i+\sum_j\frac{\partial{\mathcal L}}{\partial u_{j,i}}
 \left(\eta_j-\sum_k u_{j,k}\theta_k\right)-f_i,
\label{eq25}
\end{equation}
which satisfies the conservation law
\begin{equation}
\sum_i\frac{d I_i}{d x_i}=0.
\label{eq26}
\end{equation}

\subsection{Reduction of order of an ODE}

Let us consider an ODE $x^{(k)}=f(t,x,x',\ldots,x^{(k-1)})$ where $x^{(k)}$ is the k-th derivative of $x$ with
respect to $t$, admitting a symmetry generator ${\bf G}_1$. Using the canonical coordinates
$r$ (dependent variable) and $s$ (independent variables), we have ${\bf G}_1=\partial/\partial r$, and
consequently the original equation is cast in the form
$r^{(k)}=g(s,r',\ldots,r^{(k-1)})$, for some function $g$, which is an ODE of order $k-1$ in $u=r'$.

Now suppose the original equation admits another symmetry generator
${\bf G}_2$. It can also be rewritten using the same canonical variables. 
If the extended generator ${\bf G}_2^{(k)}$ is such that it does not act on $s$ directly
but only on its derivatives,
then it can be used for a further reduction of order. 
This procedure can then be iterated for any number of generators, provided that at each step
the generator used acts only in the remaining derivatives.
Of course if $m=n$ the system can be completely solved. This reduction is possible iff the Lie
algebra spanned by the $k$ generators is solvable~\cite{schwarz3}.

\subsection{Equations admitting a symmetry group}

In many situations the equations describing a given system are not known in closed form,
but some of its symmetries are known. This is the case for instance if an underlying kinematical
group (e.\ g.\ the Poincar\'e or Lorenz group) is imposed by the physics of the system.
This is frequently the case for transport equations for which no complete general theory exists~\cite{liboff}.
One may hope that using the knowledge of symmetries may determine a class of
equations for the problem at hand. In this way,
Let us consider a set of $p$ equations on the unknowns $u_i$, $i=1,\ldots,n$, of the form:
\begin{equation}
\sum_{j=1}^n \Delta_i^j u_j=0,\hspace{3mm}i=1,\ldots,p,
\label{nonlineq}
\end{equation}
where $\Delta_i^j$ is a differential operator which can be non-linear. The system~(\ref{nonlineq})
defines a class of equations if the operators $\Delta_i^j$ depend
on unknown functions. Now we impose that eq.~(\ref{nonlineq}) is invariant under
symmetry transformations generated by ${\bf G}_i$; $i=1,\ldots,k$, forming a system of differential equations
for the unknown functions in $\Delta_i^j$, that can be solved in some cases. This is implemented in
SADE in the routine {\tt equivalence}. The main shortcoming here is that the system to be solved
is non-linear.

\section{Heuristics for solving the determining system}

No proved fully general, finite and terminating algorithm exists for the solution of linear
or non-linear over-determined systems of partial differential equations, of the form obtained
as determining systems for Lie, Lie-B\"acklund and nonclassical symmetries. Here we present the heuristics used in
the present package. For linear systems, it was able to efficiently solve all test cases, spanning a large number
of equations ranging from the simple harmonic oscillator to equations describing coupled relativistic fields.

\subsection{Over-determined system of linear partial differential equations}

The basic idea is to solve simpler equations first, always trying to simplify further the system.
Of course the meaning of ``simpler'' is quite subjective and our definition will become clear below.
In a few cases it is necessary to
append the determining system with integrability conditions for some, or all, of its equations. This is done using the MAPLE built-in
routine rifsimp, when it is applicable, or otherwise using a slightly modified version of the Kolchin-Ritt algorithm
with sorting~\cite{mansfield2}. The reduction to the involutive form (see~\cite{rif,rif2} for a proper definition)
is usually very expensive in computational
time, and should be done only if the system cannot be solved otherwise, and after some preliminary simplifications.
Another strategy is to reduce only a subsystem with equations containing up to a prescribed number of terms.

In what follows the number of terms in an equation is the number of its summands. Parameters
controlling the flow of the solution algorithm are specified in global variables that can
be modified by the user, and with default values given below.
These are the main steps of our algorithm:

\begin{enumerate}

\item Solve all algebraic equations with a maximum of 2 terms\label{primeiro}.

\item Solve all differential equations of the form\label{singlediff}
\begin{equation}
\frac{\partial^k f}{\partial x_{i_1}\ldots x_{i_k}}=0,
\end{equation}
where $f$ is any unknown in the determining system.

\item Solve any algebraic equation in the original unknowns $\theta_i$ and $\eta_i$.

\item Reduce to involutive form the subset of equations with at most $N_1$ terms.

\item \label{lidecomp} If any equation can be written as an expansion in linearly independent functions, then
equate each coefficient to zero. Repeat until no more such decomposition is possible.

\item Repeat step~\ref{singlediff}. If any equations is solved, go to step~\ref{resolve2}.

\item Completely reduce to involutive form, and in case it succeeds, go to step~\ref{resolve2}.

\item \label{diffsolve} Solve all equations with at most $N_2$ terms that can integrated as ODE's in one of the
unknowns. If no equation can be solved, then repeat with $N_2+3$ terms, and so on
up to the maximal value $N_3$. If any equation is solved, then go to step~\ref{resolve2}.

\item Solve one ODE with any number of terms. If it succeeds, go to step~\ref{primeiro}.

\item \label{resolve2} Repeat step~\ref{primeiro}.

\item Look for all equations that are expansions in linearly independent functions, and equate to zero each
coefficient of the expansion.

\item Repeat step~\ref{primeiro}.

\item Repeat step~\ref{singlediff}.

\item Repeat step~\ref{diffsolve}.

\item Solve all algebraic equations (with any number of terms). If any equation is solved, then go
to step~\ref{resolve2}.

\item Reduce to the involutive form. If succeeds, go to to step~\ref{resolve2}.

\item Look for one equation of the form $f_1(x_1,x_2)=f_2(x_1,x_3)$ and replace it by
$f_1(x_1,x_2)=f_3(x_1)$ and $f_2(x_1,x_3)=f_3(x_1)$, where $f_1$ and $f_2$ are two unknowns 
in the system and $f_3$ a new unknown. If succeeds go to step~\ref{resolve2}.

\item Repeat step~\ref{diffsolve}.

\end{enumerate}
The whole algorithm is repeated until the system is completely solved or no additional simplification
occurs. The following global variables are used:
\begin{eqnarray*}
& & N_1={\tt SADE[partial\_reduction]}\\
& & N_2={\tt SADE[\_ne]}\\
& & N_3={\tt SADE[\_nmaxeq]}
\end{eqnarray*}
with default values $N_1=N_2=5$ and $N_3=8$.

\subsection{Non-linear systems}

Solving non-linear overdetermined systems of PDE's, as those resulting from the determination of
nonclassical symmetries, is a very difficult task. Its implementation in SADE is still under development,
but can nevertheless be used in some interesting cases. There are other more efficient algorithms, such as the one used
in the REDUCE package CRACK~\cite{wolf1}. Our algorithm can be roughly sketched as:

\begin{enumerate}

\item Solve all linear equations\label{um}.

\item Reduce the resulting system to the involutive form.

\item Solve all linear equations.

\item Decomposes equations which are expansions in linearly independent functions\label{li}.

\item Solve a single nonlinear ODE. If any equation is solved, repeat step~\ref{li} and go to step~\ref{um}.

\item Solve all purely algebraic equations, considering multiple solutions.
If any equation is solved, then go to step~\ref{um}.

\end{enumerate}
There are also options for reducing the determining system to involutive form before trying to solve it,
and to use the MAPLE builtin routine for solving PDE's.

\section{Package Commands}

Here we briefly describe each command available in SADE. The examples given in
section~\ref{illexamples} should be self explanatory and complementary to this section. The following
abbreviations are used for the input arguments:
\begin{description}
\item {\tt\bf eqs}: a single or a set of differential equations.
\item {\tt\bf unks}: the unknowns in {\tt eqs}.
\item {\tt\bf gen}: a symmetry generator, written in SADE notation (see section~\ref{illexamples}).
\item {\tt\bf depvars}: list of dependent variables.
\item {\tt\bf indepvars}: list of independent variables.
\item {\tt\bf vars}: list of dependent and independent variables.
\item {\tt\bf der\_order}: list with the derivative order of each dependent variable in the independent variables (see section~\ref{casinvsec}).
\item {\tt\bf drvs}: set with derivatives of unknowns in {\tt eqs}.
\item {\tt\bf subs\_rule}: a substitution rule.
\item {\tt\bf funcs}: a set with new undetermined functions.
\item {\tt\bf name}: a maple variable name.
\item {\tt\bf opt}: optional arguments.
\item {\tt\bf determining}: optional argument to return only the determining system;
\item {\tt\bf involutive}: optional argument to reduce the determining system to involutive form.
\item {\tt\bf params}: a set on free parameters.
\end{description}

Package commands and corresponding inputs are given in the following listing:

\begin{description}
\item {\tt\bf liesymmetries(eqs,unks,opt)}:
Computes Lie symmetry generators.
Optional arguments: {\tt determining}, {\tt involutive},
{\tt case=$\:n$}, $n$ integer, only the
case with $\theta_n=1$, $\theta_i=0$ ($i<n$) is considered.
{\tt builtin} - solves the determining system using the MAPLE builtin command {\tt pdsolve}.\\
{\tt default\_parameters} - the determining system is solved using default parameters
reducing CPU time, although the system may not be completely solved.\\
\item {\tt\bf ncsymmetries(eqs,unks,opts)}:
Computes nonclassical symmetry generators of DE's.
Optional arguments: {\tt determining} and {\tt involutive}.\\
\item {\tt\bf LBsymmetries(eqs,unks,opts)}: Obtains Lie-B\"acklund symmetry generators.
Optional arguments: {\tt determining}, {\tt involutive} and\\ {\tt parameter = paramset}
- computes the generators with conditions on the free parameters in {\tt paramset}.\\
\item {\tt\bf lindsolve(eqs,unks)}: Solves a linear overdetermined system of PDE's.\\
\item {\tt\bf nonlindsolve(eqs,unks)}: Solves a non-linear overdetermined system of partial differential equations.\\
\item {\tt\bf casimir\_invariant(\{$\tt\bf gen_1,gen_2,\ldots$\},depvars,indepvars,der\_order)}:\\
Computes the Casimir invariants of a set of generators.\\
\item {\tt\bf ansatz(subs\_rule,funcs)}: Applies a set of ans\"atze to the determining equations. This routine can be used either once the determining
system is obtained or if SADE could not completely solve the determining system. The elements of {\tt subs\_rule}
must be given in the form {\tt function = expression} with new undetermined functions in {\tt expression} specified in {\tt funcs}.\\
\item {\tt\bf noether(lagrangian,funcs,gen)}: Computes the N\"other conserved currents or first-integrals from a lagrangian function.\\
\item {\tt\bf equivalence(eqs,\{$\tt\bf gen_1,gen_2,\ldots$\},funcs)}: Obtains the most generic form of a class of equations
admitting a symmetry algebra.\\
\item {\tt\bf comm($\tt\bf gen_1,gen_2$,vars)}: Commutator of two linear operators (generators).\\
\item {\tt\bf com\_table(\{$\tt\bf gen_1,gen_2,\ldots$\},vars,name)}: Commutation table of a set of infinitesimal
generators. {\tt name} specifies a name to represent each generator.\\
\item {\tt\bf AdjointRep(\{$\tt\bf gen_1,gen_2\ldots$\},vars,name,parameter)}: Computes the table with the action of adjoint maps on each generator of a Lie Algebra.
{\tt name} is used to represent each generator in the table and
{\tt parameter} specifies the adjoint Lie group parameter\\
\item {\tt\bf StructConst(\{$\tt\bf gen_1,gen_2\ldots$\},vars)}: Computes an array with the structure constants of a Lie algebra.\\
\item {\tt\bf linear\_rep(operator,\{$\tt\bf gen_1,gen_2\ldots$\},vars,name)}: Determines the most general linear operator representing a class of differential
equations defined by {\tt operator} admitting a symmetry algebra.\\
\item {\tt\bf PDEreduction(eqs,unks,\{$\tt\bf gen_1,gen_2\ldots$\})}: Obtains the reduced form of a PDE or a PDE system from a set of symmetry generators.
For a system with $M$ independent variables $K$ symmetry generators can be used to reduce to a new system depends with $M-K$ (transformed) independent variables.\\
\item {\tt\bf invariant\_sol(eqs,unks,\{$\tt\bf gen_1,gen_2\ldots$\})}: Obtains invariant solutions of a PDE or a system of PDE's using symmetry generators.\\
\item {\tt\bf issolvable(\{$\tt\bf gen_1,gen_2\ldots$\},vars)}: Tests if a Lie algebra is solvable.\\
\item {\tt\bf canonical\_basis(\{$\tt\bf gen_1,gen_2\ldots$\},vars)}: Computes the canonical basis of a Lie algebra.\\
\item {\tt\bf derived\_subalg(\{$\tt\bf gen_1,gen_2\ldots$\},vars)}: Computes the generators of the derived subalgebra of a Lie algebra.\\
\item {\tt\bf odesolver(eqs,\{$\tt\bf gen_1,gen_2\ldots$\},unks,opt)}: Solves an ODE by successive reductions using a solvable Lie algebra.
Optional argument: {\tt transformation} - returns only the transformation of variables solving the system.\\
\item {\tt\bf reduce\_ode\_sist(eqs,\{$\tt\bf gen_1,gen_2\ldots$\},depvars${}_1$,indepvar${}_1$,depvars${}_2$,\\ indepvar${}_2$)}: Reduces by one the dimension of a system of first order ODE's
using a symmetry generator. Note that new dependent and independent variables must be given and are represented by the index 2. Index 1 denotes original variables.\\
\item {\tt\bf ode\_reduce\_order1(eqs,gen,depvar${}_1$,indepvar${}_1$,depvar${}_2$,indepvar${}_2$)}:\\ Reduces by one the order a a single ODE using a symmetry generator.\\
\item {\tt\bf ode\_invsolution(eqs,unks,gen)}: Obtains invariant solutions for a single ODE.\\
\item {\tt\bf conserved(eqs,unks,params,n,opt)}: Obtains the QP-invariants of a QP first order system by computing a Darboux polynomial up to degree {\tt n}.
Optional arguments: {\tt Groebner} - a Gr\"obner basis  computation is used to solve the polynomial system of determining  equations.
{\tt positive} - the results are simplified to the positive orthant.
{\tt surfaces} - returns the defining equations for invariant hyper-surfaces.\\
\item {\tt\bf QPsymmetries(eqs,unks,params,n)}: Determines QP symmetry generators with $n$ the degree of the polynomial quasi-symmetry~\cite{qpsi}.\\
\item {\tt\bf verif\_if\_inv(eqs,unks,params)}: Determines parameter values such that non-trivial QP first-integrals
(i.\ e.\ with non-integer exponents) may exist.\\
\item {\tt\bf constlog(eqs,unks,params,n)}: Computes first-integrals of the form
$P_1(x)+\log(x^\xi)$ and $P_2(x,\ln(x))$,
with $P_1$ and $P_2$ polynomials of degree {\tt n} and $x$ stands for all dependent variables.
\end{description}

\section{Illustrative examples}
\label{illexamples}

We present some illustrative cases of a basic use of SADE.
A whole suite of examples, including more complex problems, is given with the distribution files.

\subsection{Lie symmetries}

Let us consider first as a simple example the heat diffusion equation: $\partial u/\partial t=\partial^2 u/\partial x^2$.
The input to determine its Lie symmetries and the corresponding output are:
\begin{flushleft}
$
\tt >g:=liesymmetries(\{diff(u(x,t),t)=diff(u(x,t),x,x)\},[u(x,t)]);
$
\end{flushleft}
\begin{eqnarray*}
\sf \displaystyle g := & & \left[ \left\{ D_{{t}},D_{{x}},uD_{{u}},{\it \_F1} \left( t,x \right) D_{{u}},2\,tD_{{t}}+xD_{{x}},uxD_{{u}}-2\,tD_{{x}}
\mbox{},\right.\right.\\
 & & \left.\left.{t}^{2}D_{{t}}+1/4\, \left( -2\,ut-u{x}^{2} \right) D_{{u}}+txD_{{x}} \right\},\right. \\
 & & \left. \left\{ {\frac {\partial ^{2}}{\partial {x}^{2}}}{\it \_F1} \left( t,x \right) -{\frac {\partial }{\partial t}}{\it \_F1} \left( t,x \right)  \right\} \right]
\end{eqnarray*}
The first element of the list in the output is the set of infinitesimal symmetry generators. The second element
is a list of constraints on these generators, which in this case is the original equation.
The element $\sf D[\alpha]\rightarrow D_\alpha$ in the output
stands for the directional derivative $\partial/\partial \alpha$, where $\alpha$ is any of the dependent or independent
variables. The commutator of two generators is computed as:
\begin{flushleft}
$
\tt >comm(D[x],-1/2*u*x*D[u]+t*D[x],[u,x,t]);
$
\end{flushleft}
$$
\sf -\frac{1}{2}~u ~{\rm D}_{u }
$$
The complete commutation relations for the finite dimensional algebra can be computed as follows:
\begin{flushleft}
$
\tt >gen:=convert(g[1]\,\, minus\,\, \{\_F1(t,x)*D[u]\},list);
$

$
\tt >com\_table(gen,[u,x,t],G);
$
\end{flushleft}

$$
\left[ \begin {array}{cccccc}
 0&0&0&2\,{\it G1}&-2\,{\it G2}&-1/2\,{\it G3}+{\it G4}\\
 \noalign{\medskip}0&0&0&{\it G2}&{\it G3}&-1/2\,{\it G5}\\
 \noalign{\medskip}0&0&0&0&0&0\\
 \noalign{\medskip}-2\,{\it G1}&-{\it G2}&0&0&{\it G5}&2\,{\it G6}\\
 \noalign{\medskip}2\,{\it G2}&-{\it G3}&0&-{\it G5}&0&0\\
 \noalign{\medskip}1/2\,{\it G3}-{\it G4}&1/2\,{\it G5}&0&-2\,{\it G6}&0&0\end {array} \right]
$$

\subsection{Lie-B\"acklund symmetries}

To illustrate how dependencies on derivatives of dependent variables are handled when computing Lie-B\"acklund symmetries,
we consider the following two-dimensional PDE system~\cite{steeb}:
\begin{equation}
\frac{\partial^2 u_2}{\partial x_1^2}=\frac{1}{2}\frac{\partial u_2}{\partial x_2};\hspace{5mm}
\frac{\partial^2 u_1}{\partial x_1^2}=\frac{\partial u_1}{\partial x_2}-\frac{u_2^2}{2}.
\label{exsteebortho}
\end{equation}
Requiring, for instance, that the evolutionary form of the symmetry generators depends on
$\partial u_1/\partial x_1$, $\partial^2 u_1/\partial x_1^2$,
$\partial u_2/\partial x_1$ and $\partial^3 u_2/\partial x_1^3$,
implies that the independent derivatives are $\partial u_1/\partial x_1$, $\partial u_1/\partial x_2$, $\partial u_2/\partial x_1$
and $\partial^2 u_2/\partial x_1\partial x_2$. Here are the corresponding input and output:
\begin{eqnarray*}
\tt > & & LBsymmetries(eqs,[u1(x1,x2),u2(x1,x2)],\\
 & & \{diff(u1(x1,x2),x1),diff(u1(x1,x2),x1\$2),diff(u2(x1,x2),x1),\\
 & & diff(u2(x1,x2),x1\$3)\});
 \end{eqnarray*}
 \begin{eqnarray*}
 & & \left[\left\{{-1/2\,{\frac { \left( -2\, \left( {\frac {\dd}{\dd{\it x1}}}{\it u2} \right) {\it u2}+{\frac {\dd^{2}}{\dd{\it x1}d{\it x2}}}{\it u2} \right) D_{{{\it u1}}}
\mbox{}}{{\it u2}}}+1/2\,{\frac { \left( {\frac {\dd^{2}}{\dd{\it x1}\dd{\it x2}}}{\it u2} \right) D_{{{\it u2}}}}{{\it u2}}}},\right.\right.\\
 & & 1/2\,{\frac { \left( -{\frac {\partial }{\partial {\it x2}}}{\it \_F1} \left( {\it x1},{\it x2} \right) +2\,{\it \_F1}
\left( {\it x1},{\it x2} \right) {\it u2} \right) D_{{{\it u1}}}}{{\it u2}}}\\
& & \left.+1/2\,{\frac { \left( {\frac {\partial }{\partial {\it x2}}}{\it \_F1}\left( {\it x1},{\it x2} \right)  \right) D_{{{\it u2}}}}{{\it u2}}}\right\},\\
& & \left\{ 2\,{\frac {\partial ^{2}}{\partial {{\it x1}}^{2}}}{\it \_F1} \left( {\it x1},{\it x2} \right)
-{\frac {\partial }{\partial {\it x2}}}{\it \_F1} \left( {\it x1},{\it x2} \right)  \right\},\\
& & \left.\left\{ {\frac {\dd}{\dd{\it x1}}}{\it u1},{\frac {\dd}{\dd{\it x2}}}{\it u1},
{\frac {\dd}{\dd{\it x1}}}{\it u2},{\frac {\dd^{2}}{\dd{\it x1}\dd{\it x2}}}{\it u2} \right\}\right]
\end{eqnarray*}
The first element is a set with the Lie-B\"acklund symmetry generators, the second element a set with constraints on the generators (an empty set if none), and the
third element the set of independent derivatives.

\subsection{Solving an ODE}

Let us consider the equation~\cite{hydon}:
\begin{equation}
\left(\frac{d\,u}{dx}\right)^5\frac{d^3u}{dx^3}-3\left(\frac{d\,u}{dx}\right)^4\left(\frac{d^2u}{dx^2}\right)^2-\left(\frac{d^2u}{dx^2}\right)^3=0,
\label{lesssimp}
\end{equation}
with $u=u(x)$. First compute its Lie symmetries:
\begin{eqnarray*}
\sf > eq:= & & \sf diff(u(x),x)\symbol{94}5*diff(u(x),x,x,x)-diff(u(x),x,x)\symbol{94}3\\
 & & \sf -3*diff(u(x),x)\symbol{94}4*diff(u(x),x,x)\symbol{94}2:\\
& & \sf liesymmetries(eq,[u(x)]);
\end{eqnarray*}
$$
\it [ \left\{ D_{{u}},D_{{x}},uD_{{x}},uD_{{u}}+3/2\,xD_{{x}} \right\} , \left\{  \right\} ]
$$
then chose a solvable subalgebra:
\begin{eqnarray*}
 > & & \sf ls:= \left\{ D_{{u}},D_{{x}},uD_{{x}} \right\}:\\
 & & \sf issolvable(ls,[u,x]);\\
\end{eqnarray*}
$$
\it true
$$

Here are the inputs for computing the solutions and displaying the first one (there are three branches):
\begin{eqnarray*}
 & & > \sf sol:=odesolver(eq,\{D[u],D[x],u*D[x]\},[u(x)]):\\
 & &  >  \sf sol[1];\\
 & & u ( x ) =1/2\,{\it \_C1}-1/2\, ( 1/2\,(-12\,x
+6\,{\it \_C2}\,{\it \_C1}-12\,{\it \_C3}-{{\it \_C2}}^{3}\\
& & +2\,(
36\,{x}^{2}-36\,x{\it \_C2}\,{\it \_C1}+72\,x{\it \_C3}+6\,x{{\it \_C2
}}^{3}\\
& & +9\,{{\it \_C2}}^{2}{{\it \_C1}}^{2}-36\,{\it \_C2}\,{\it \_C1}
\,{\it \_C3}-3\,{{\it \_C2}}^{4}{\it \_C1}+36\,{{\it \_C3}}^{2}\\
& & 
+6\,{\it \_C3}\,{\it \_C2}^{3})^{1/2})^{1/3} +1/2\,{\it \_C2}^{2}/((
-12\,x+6\,{\it \_C2}\,{\it \_C1}-12\,{\it \_C3}\\
& & 
-{\it \_C2}^{3}+2\,
(36\,{x}^{2}-36\,x{\it \_C2}\,{\it \_C1}+72\,x{\it \_C3}+6\,x{
\it \_C2}^{3}\\
 & & +9\,{\it \_C2}^{2}{\it \_C1}^{2}-36\,{\it \_C2}\,{
\it \_C1}\,{\it \_C3}-3\,{\it \_C2}^{4}{\it \_C1}+36\,{\it \_C3}^{2}\\
 & & +6\,{\it \_C3}\,{\it \_C2}^{3})^{1/2})^{1/3})1/2\,{\it \_C2} )^{2}
\end{eqnarray*}

\subsection{Invariant solutions}

The Burgers equation in one dimension can be rewritten as a set of two one-dimensional equations by defining $v= \partial u/\partial x$~\cite{burgers}:
\begin{equation}
{\frac {\partial u}{\partial x}}-v=0;\hspace{5mm}{\frac{\partial u}{\partial t}}
+uv-{\frac {\partial v}{\partial x}}=0.
\end{equation}
Since the number of independent variables is two, only a single symmetry generator is necessary to obtain a group invariant solution.
In this form, one of the symmetry generators admitted by the Burgers equation is:
\begin{equation}
{\bf G}= \left( ut-x \right) D_{{u}}+ \left( 2\,vt-1 \right) D_{{v}}-{t}^{2}D_{{t}}-txD_{{x}}.
\end{equation}
The associated invariant solution is obtained using the following input in SADE:
\begin{eqnarray*}
 > & &  \sf invariant\_sol(eq,[u(x,t),v(x,t)],\{(u*t-x)*D[u]\\
 & & \sf +(2*v*t-1)*D[v]-t^2*D[t]-t*x*D[x]\});\\
\end{eqnarray*}
\begin{eqnarray*}
 & & \hspace{-10mm}\{  \{ u ( x,t ) =- ( -{\it \_C1}\,x+\tanh ( 1/2\,{\frac {x+{\it \_C2}\,t}{{\it \_C1}\,t}} )  ) {{\it \_C1}}^{-1}{t}^{-1},\\
 & & \hspace{-10mm}v(x,t)=1/2\, ( 2\,t ( \cosh ( 1/2\,{\frac {x+{\it \_C2}\,t}{{\it \_C1}\,t}} )  ) ^{2}{{\it \_C1}}^{2}-1 ) {t}^{-2}\\
 & & \hspace{-10mm}
( \cosh ( 1/2\,{\frac {x+{\it \_C2}\,t}{{\it \_C1}\,t}} )  ) ^{-2}{{\it \_C1}}^{-2} \}  \}
\end{eqnarray*}

\subsection{PDE reduction}

Let us consider a massless nonlinear Klein-Gordon equation with a $\lambda\phi^4/4$ self-interaction potential:
\begin{eqnarray*}
  > & & \sf p:=phi(x,y,z,t):\\
 & & \sf eq:=diff(p,x,x)+diff(p,y,y)+diff(p,z,z)-diff(p,t,t)+lambda*p^3:
\end{eqnarray*}
This equation can be reduced to a PDE with two independent variables using the symmetry generators:
\begin{equation}
{\bf G}_1  =  yD_{{t}}+tD_{{y}},\hspace{5mm}
{\bf G}_2  =  zD_{{x}}-xD_{{z}},
\end{equation}
a Lorenz boost in the $y$ direction and a spatial rotation around the $y$ axis, respectively.
The reduced equation is obtained using the SADE command:
\begin{eqnarray*}
 > & & \sf PDEreduction(eq,[phi(x,y,z,t)],{yD_{t}+tD_{y},zD_{x}-xD_{z}});
\end{eqnarray*}
\begin{eqnarray*}
& &
\{ [ \{ \phi ( x,y,z,t ) ={\it \_F1} ( {x}^
{2}+{z}^{2},-{y}^{2}+{t}^{2} )  \} ,\\
& & 
 \{ 4\, ( {
\frac {\partial ^{2}}{\partial {{\it \xi1}}^{2}}}{\it \_F1} ( {
\it \xi1},{\it \xi2} )  ) {\it \xi1}+4\,{\frac {\partial }{
\partial {\it \xi1}}}{\it \_F1} ( {\it \xi1},{\it \xi2} ) -4\,
{\frac {\partial }{\partial {\it \xi2}}}{\it \_F1} ( {\it \xi1},{
\it \xi2} ) \\
& & -4\, ( {\frac {\partial ^{2}}{\partial {{\it \xi2
}}^{2}}}{\it \_F1} ( {\it \xi1},{\it \xi2} )  ) {\it 
\xi2}+\lambda\, ( {\it \_F1} ( {\it \xi1},{\it \xi2} ) 
 ) ^{3} \} , \\
& & \{ {\it \xi1}={x}^{2}+{z}^{2},{\it \xi2}=-
{y}^{2}+{t}^{2} \} ] \} 
\end{eqnarray*}
The output is a set with the different possible reductions (one in the present case). Each element is a list:
the first element defines the relation between the original unknown(s) and the reduced dependent variable(s).
The second element is the set of reduced equations, and the last element is a set with the similarity
variables $\xi_i$  used to reduce the original equation(s).

\subsection{N\"other theorem}

As an example, we consider the relativistic massless scalar field in 1+1 dimensions, with
lagrangian density:
\begin{equation}
{\cal L}=\frac{\partial^2\phi}{\partial x^2}-\frac{\partial^2\phi}{\partial t^2}=0.
\label{scalarlag}
\end{equation}
In this case a conserved current $\bf I$ is a two-vector function of $\phi$, $x$ and $t$ such that
$\partial I_1/\partial x+\partial I_2/\partial t=0$.
They can be obtained in the following way:
\begin{flushleft}
$
\tt >noether(diff(phi(x,t),x)\symbol{94}2/2-diff(phi(x,t),t)\symbol{94}2/2,[phi(x,t)]);
$
\end{flushleft}
\begin{eqnarray*}
& &\sf \left\{ \left[1/2\, \left( {\frac {\dd}{\dd x}}\varphi  \right) ^{2}+1/2
\, \left( {\frac {\dd}{\dd t}}\varphi  \right) ^{2},
- \left( {\frac {\dd}{\dd t}}\varphi  \right) {\frac {\dd }{\dd x}}\varphi \right],\left[-2\,{\frac {\dd}{\dd t}}\varphi ,2\,{\frac {\dd }{\dd x}}
\varphi \right],\right.\\
& &\sf\left.
\left[ \left( {\frac {\dd}{\dd t}}\varphi  \right) {\frac {\dd}{\dd x}}\varphi ,-1/2\, \left( {\frac {\dd}{\dd x}}\varphi  \right) ^{2}
-1/2\, \left( {\frac {\dd}{\dd t}}\varphi  \right) ^{2}\right] \right\} ,\,\left[t,x\right]
\end{eqnarray*}
Here the output is a sequence. The first element is the set of conserved currents, and the second element the
independent variables defining the component ordering used for the components of ${\bf I}$.

\subsection{Equations admitting a symmetry algebra}

Let us consider the following family of PDE's:
\begin{equation}
\frac{\partial u}{\partial t}+\frac{\partial(f u)}{\partial x}+\frac{\partial^2(d u^2)}{\partial x^2}=0,
\label{geneq}
\end{equation}
where $f=f(x,t)$ and $d=d(x,t)$ are functions to be determined.  Requiring that eq.~(\ref{geneq}) admits
${\bf G}_1=\partial/\partial t$, ${\bf G}_2=\partial/\partial x$ and ${\bf G}_3=(x/2)\partial/\partial x+t\partial/\partial t$
as symmetry generators restricts the postulated generic form~(\ref{geneq}):
\begin{flushleft}
$
\sf > gen := \{D[t], D[x], (1/2)*x*D[x]+t*D[t]\}:
$
$
\sf >eq:=diff(u(x,t),t)+diff(f(x)*u(x,t),x)+diff(d(x)*u(x,t)\symbol{94}2,x,x):
$
$
\sf>equivalence(eq,[u(x,t)],gen,\{f(x,t),d(x,t)\});
$
\end{flushleft}

\begin{eqnarray*}
 & &\{ [{\frac {\partial }{\partial t}}u ( x,t ) +2\,{\it \_C1}\, ( {\frac {\partial }{\partial x}}u ( x,t )  ) ^{2}+2\,
{\it \_C1}\,u ( x,t ) {\frac {\partial ^{2}}{\partial {x}^{2}}}u ( x,t ) , \{  \} ],\\
& & [{\frac {\partial }{\partial t}}u ( x,t )
+ ( {\frac {d}{dx}}f ( x )  ) u ( x,t ) +f ( x ) {\frac {\partial }{\partial x}}u ( x,t ) , \{  \} ] \}
\end{eqnarray*}
The output is a set of lists. In each list, the first element is a restricted form for
the class of equations, the second element is the set of remaining equations still to be solved, none in this case.

\subsection{Casimir invariants}
\label{casinvsec}

The order of the derivatives in a Casimir invariant is specified by a list such that each element is a list with
the maximum derivative order of the corresponding dependet variable in each independent variable. 
Note that the order of the derivatives in the invariants are one order higher than specified in the input
(this is due to the specific algorithm used in the computation):

\begin{eqnarray*}
\sf  & > &  casimir\_invariant({x*D[t]+t*D[x],v*D[v]+u*D[u]-x*D[x]},\\
 & & \hspace{20mm}[u],[x,t],[[2,2],[2,2]]);
\end{eqnarray*}
\begin{eqnarray*}
 & & \left[-\frac{\left(\frac{\partial ^{2}}{\partial x ^{2}}~u \left(x ,t \right)\right)~\left(\frac{\partial ^{2}}{\partial t ^{2}}~u
\left(x ,t \right)\right)-\left(\frac{\partial ^{2}}{\partial x ~\partial t }~u \left(x ,t \right)\right)^{2}}{u \left(x ,t \right)^{4}},\right.\\
 & & \frac{\partial }{\partial x }~u \left(x ,t \right),\frac{\partial }{\partial t }~u \left(x ,t \right),
\frac{\partial ^{3}}{\partial x ^{3}}~u\left(x ,t \right),
\frac{\partial ^{3}}{\partial t ^{3}}~u \left(x ,t \right),\\
 & & \left.\frac{2~x ~t ~\left(\frac{\partial ^{2}}{\partial x ~\partial t }~u \left(x ,t \right)\right)
+x^{2}~\left(\frac{\partial ^{2}}{\partial x ^{2}}~u \left(x ,t \right)\right)+t ^{2}~\left(\frac{\partial ^{2}}
{\partial t ^{2}}~u \left(x ,t \right)\right)}{u \left(x ,t \right)}\right]
\end{eqnarray*}

\section{Lie Symmetries Benchmarks}

\begin{table}
\caption{Benchmarks for Lie Symmetry determination.}
\begin{tabular}{lrrrr}
\hline
\hline
System & $N_{eq}$ & SADE & DESOLV & PDEtools \\ \hline
Heat equation~\cite{olver} & 9 & 0.4 & 0.13 & 0.4 \\
%Vaidya system~\cite{ibragimov2} & 20 & 9.0 & 81.8 & 22.3 \\
Klein-Gordon~\cite{ibragimov2} & 32 & 1.1 & 0.69 & 0.9\\
Magneto-Hydro-Dynamics~\cite{ibragimov2} & 39 & 14.4 & 12.9 & 13.4 \\
Navier-Stokes~\cite{kumei} & 125 & 3.9 & 4.8 & 2.9 \\
Dirac~\cite{steeb} & 352 & 14.7 & 20.1 & --- \\
Gross-Neveu~\cite{gross} & 352 & 15.4 & 211 & --- \\
Maxwell-Dirac~\cite{steeb} & 2,621 & 27.8 & 1,121. & --- \\
Yang-Mills SU(2)~\cite{schwarz2} & 12,361 & 82 & 188& --- \\
Yang-Mills SU(3)~\cite{sun} & 175,042 & 13,847 & --- & --- \\
\hline
\end{tabular}
\\
\label{tabela2}
\end{table}

It is beyond the scope of the present work to do an exhaustive comparison of SADE to other similar packages
(see References~\cite{butcher,hereman} for an assessment for previous packages).
We present only a comparison for the computation of Lie symmetries (the core of SADE), and leave a more thorough
comparison to be presented in a future work. Since the release 11 of MAPLE
it became possible to compute Lie symmetries using its native package PDEtools. The package GEM relies on the solution
of the determining system using MAPLE built-in routines, leading to large computational effort in CPU time and memory
when solving the determining system, becoming
intractable for systems such as Maxwell-Dirac or SU(2) and SU(3) Yang-Mills field equations.
In this way, and considering only MAPLE packages, table~\ref{tabela2} shows CPU
times for the determination of Lie symmetries (solving the determining system and obtain the
symmetry generators in explicit form) of some representative systems using SADE, DESOLV and PDEtools. 
All computations were performed on a i5 2.40 GHz computer, and using the corresponding
automated routines for computing Lie symmetries
in each package: {\it symmetry} and {\it genvec} in DESOLV, {\it liesymmetries} in SADE and {\it Infinitesimals} in PDEtools.
The absence of a value in the table means that either the computation
was not completed after a very long time,
or that the package was unable to obtain all symmetries of the equation.
From table~\ref{tabela2} we see that PDEtools handles only simpler systems (the three first lines in the table),
with DESOLV being the fastest and PDEtools the slowest. Nevertheless, for the remaining cases in table~\ref{tabela2},
with a number of equations in the determining system ranging from a few hundreds to hundreds of thousands, SADE
performs better than DESOLV. For the Yang-Mills $SU(3)$ field equations, SADE took a little less than $6$ hours while
DESOLV produced no output after more than $60$ hours. On its turn, PDEtools missed some symmetries of the Navier-Stokes and all of the Dirac equations, and for the
remaining equations was unable to terminate the computations after a considerable long time.
We used MAPLE 14 for timings, except for DESOLV which performs better in MALE 13 (for the latest version available to the authors).

\section{Concluding Remarks and Perspectives}

The present package implements symmetry methods for differential equations in MAPLE, including
Lie, nonclassical, Lie-B\"acklund and potential symmetries, the Quasi-Polynomial formalism and the computation of invariant solutions and
reduction of ODE's and PDE's.
SADE is well suited to handle more complicated systems,
such as Maxwell-Dirac and Yang-Mills $SU(2)$ and $SU(3)$ equations, for which optimization is crucial.
It also obtains all symmetries for the ``difficult'' Vaidya and Jacob-Jones systems~\cite{butcher}.
We performed no direct performance comparisons besides those for Lie symmetries determination.
The implementation of Markus algebras~\cite{markus,nos1} as a tool for the computation of quasi-polynomial
first-integrals, symmetries and invariant hyper-surfaces, and a better algorithm for solving overdetermined systems of non-linear PDE's
are currently under implementation.
We also hope to implement a more complete set of routines for the computation of conservation laws for PDE's.

\section{Acknowledgments}

This work was partially supported by CNPq and CAPES (Brazilian Agencies). The authors
would like to thank G.~Grebot, A.~E.~Santana, A.~R.~Queiroz and L.~Brenig for many fruitful discussions.

\end{document}